\documentclass[twocolumn,preprintnumbers,showpacs,pre,floatfix,amsmath,amssymb,notitlepage,superscriptaddress]{revtex4-1}
\usepackage{epsfig}
\usepackage{subfigure}
\usepackage{amsmath}
\usepackage{color}
\usepackage{amssymb}
\usepackage{setspace}
\usepackage{graphicx}
\usepackage{dcolumn}
\usepackage{bm}
\usepackage{times}
\usepackage{enumerate}
\usepackage{footnote}
\input epsf

\graphicspath{{figures/}}

\begin{document}

\author{Per Sebastian Skardal}
\email{skardals@gmail.com} 
\affiliation{Department of Mathematics, Trinity College, Hartford, CT 06106, USA}
\affiliation{Departament d'Enginyeria Informatica i Matem\'{a}tiques, Universitat Rovira i Virgili, 43007 Tarragona, Spain}

\author{Juan G. Restrepo}
\affiliation{Department of Applied Mathematics, University of Colorado, Boulder 80309, Colorado, USA}

\author{Edward Ott}
\affiliation{Institute for Research in Electronics and Applied Physics, University of Maryland, College Park, Maryland 20742, USA}

\title{Frequency assortativity can induce chaos in oscillator networks}

\begin{abstract}
We investigate the effect of preferentially connecting oscillators with similar frequency to each other in networks of coupled phase oscillators (i.e., frequency assortativity). Using the network Kuramoto model as an example, we find that frequency assortativity can induce chaos in the macroscopic dynamics. By applying a mean-field approximation in combination with the dimension reduction method of Ott and Antonsen, we show that the dynamics can be described by a low dimensional system of equations. We use the reduced system to characterize the macroscopic chaos using Lyapunov exponents, bifurcation diagrams, and time-delay embeddings. Finally, we show that the emergence of chaos stems from the formation of multiple groups of synchronized oscillators, i.e., meta-oscillators.
\end{abstract}

\pacs{05.45.Xt, 89.75.Hc}

\maketitle
The synchronization of network-coupled dynamical systems~\cite{Strogatz2003,Arenas2008PR} plays a key role in many natural phenomena~\cite{Buck2988QRB,Glass1988} and engineering applications~\cite{Motter2013NaturePhysics,Strogatz2005Nature}. An important example is networks of coupled oscillators. Kuramoto showed~\cite{Kuramoto1984} that under suitable conditions, the analysis of an ensemble of $N$ oscillators can be reduced to the dynamics of phase angles for the oscillators, where oscillator $i$ has phase angle $\theta_i$ for $i=1,\dots,N$. When the oscillators are coupled by a network, the corresponding model is given by
\begin{align}
\dot{\theta}_i=\omega_i+K\sum_{j=1}^NA_{ij}\sin\left(\theta_j-\theta_i\right),\label{eq:Kuramoto}
\end{align}
where $\omega_i$ is the natural frequency of oscillator $i$, $K\ge0$ is the global coupling strength, and $\left[A_{ij}\right]$ is the network adjacency matrix that encodes the network structure ($A_{ij}=1$ if there is a network link from node $j$ to node $i$ and $A_{ij}=0$ otherwise). 

The dynamics of Eq.~(\ref{eq:Kuramoto}) and its many extensions have been the subject of a great deal of research (e.g., Refs.~\cite{Ichinomiya2004PRE,Restrepo2005PRE,Oh2005PRE,Arenas2006PRL,Barlev2011Chaos}). Recently an advance in the analysis of such systems was obtained~\cite{Ott2008Chaos,Ott2009Chaos} which posits an ansatz for the long time asymptotic form of the solution of such systems and results in a dimensionality reduction whereby the $N$--dimensional dynamics of Eq.~(\ref{eq:Kuramoto}) can be reduced to a much smaller system. This ansatz was first used on all-to-all coupled phase oscillator systems~\cite{Ott2008Chaos} (where each entry of the adjacency matrix is $A_{ij}=1$), and adapted to obtain analytical results revealing the effects of various extensions of the original Kuramoto model, including chimera states, periodic forcing, bimodal frequency distributions, time-delays, clustering, and communities~\cite{Abrams2008PRL,Childs2008Chaos,Martens2009PRE,Lee2009PRL,Skardal2011PRE,Barreto2009PRE,Skardal2012PRE}. Recently, the ansatz was extended via a mean-field technique to allow for the treatment of nontrivial network topologies~\cite{Restrepo2014EPL}, importantly shedding light on the effects of correlations between the degrees of network-connected node pairs, i.e., degree assortativity~\cite{Newman2002PRL}.

The formalism of Ref.~\cite{Restrepo2014EPL} can in principle be extended to account for assortativity based on arbitrary nodal properties, i.e., for probabilistic network generative models in which the probability that two nodes are connected is a function of preassigned nodal properties~\cite{Newman2003PRE,Pfeiffer2014}. In particular, referring to Eq.~(\ref{eq:Kuramoto}) we note that nodes are characterized not only by their in- and out-degrees ($k_i^{in}=\sum_{j}A_{ij}$, $k_{i}^{out}=\sum_{j}A_{ji}$), but also by their natural frequencies $\omega_i$. It would seem that frequency assortativity would be crucial for the dynamics of the network Kuramoto problem since cooperative interactions between pairs of connected nodes with like (unlike) frequencies would be stronger (weaker). However, so far there is no analytical means of investigating the impact of this basic consideration on network dynamics. It is the purpose of this Rapid Communication to provide and illustrate such an analytical technique for investigating this effect. Our results show that frequency assortativity can play a profound role in determining dynamical behavior. In particular, we show that frequency assortativity can induce chaos in the macroscopic system dynamics. While chaos has previously been found in the macroscopic dynamics of phase oscillator models~\cite{Alonso2011Chaos,So2011Chaos,Komarov2013PRL}, we find it remarkable that chaos and complex dynamics can arise in the simple, basic model given by Eq.~(\ref{eq:Kuramoto}) merely from frequency assortativity. In the remainder of this Rapid Communication we describe a simple model for generating networks with frequency assortativity, investigate the emergence of chaotic dynamics in such networks using numerical simulations, present a dimensionality reduction method for such networks, and finally close with a brief discussion of our results.

{\it Frequency assortativity network model.} We begin by briefly describing a model for generating oscillator networks with particular frequency-frequency correlation between neighbors, i.e., frequency assortativity. In other words, this model will allow for the construction of networks where neighboring oscillator tend to have similar or dissimilar natural frequencies. Because we wish to focus on the effect of frequency assortativity in the simplest and cleanest setting, we henceforth consider the case of an {\it undirected} network in which {\it all nodes have the same degree}. Note that, by this choice, issues of different degree distibutions, node degree-frequency correlations, and degree assortativity are, by definition, absent, thus providing an unambiguous testing ground for investigating frequency assortativity effects with no other complications. (We note that, although our subsequenct theory is for this special case, it is easily generalized to account for the other effects mentioned above.) Our model is based on the configuration model~\cite{Molloy1995} such that to each node $i=1,\dots,N$ we assign the same degree $k_i=k$. Additionally, we assign to each oscillator a {\it target frequency}, $\omega_{0,i}$ which will be used to build network connections as follows. Choosing a node $i$ that still requires at least one additional link, another node $j$ which still requires at least one link is chosen according to a probability $p_{ij}$. Each $p_{ij}$ depends on the target frequencies $\omega_{0,i}$ and $\omega_{0,j}$. In the networks used here, we use $p_{ij}\propto 0.5+c[d^\gamma/(d^\gamma+|\omega_{0,i}-\omega_{0,j}|^\gamma)-0.5]$ with $d=0.8$ and $\gamma=5$. In essence, the parameter $c$ tunes the degree of frequency assortativity: $c>0$ ($c<0$) allows oscillators to more likely make connections to other oscillators with similar (dissimilar) target frequencies, resulting in assortative (disassortative) networks. Links are made until all nodes have degree $k$. Finally, actual natural frequencies are assigned to each oscillator $i$ according to a distribution $g_{\omega_{0,i}}(\omega)$ that depends on the target frequency $\omega_{0,i}$. Here we consider the case of Lorentzian distributions
\begin{align}
g_{\omega_0}(\omega)=\frac{1}{\pi}\frac{\Delta_{\omega_0}}{(\omega-\omega_0)^2+\Delta_{\omega_0}^2},\label{eq:Model01}
\end{align}
centered at $\omega_0$ with spread $\Delta_{\omega_0}$.

\begin{figure}[t]
\centering
\epsfig{file =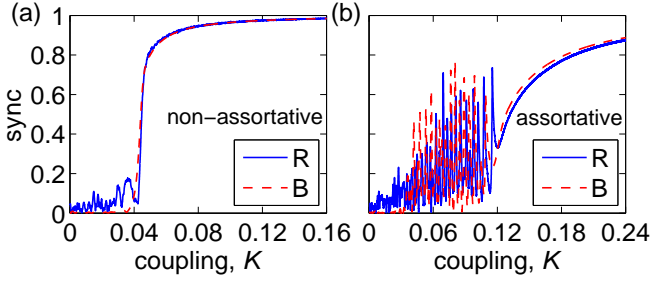, clip =,width=\linewidth }
\caption{(Color online) \textit{Synchronization in non-assortative and assortative networks.} Synchronization profiles $R$ (solid blue) and $B$ (dashed red) vs $K$  for examples of (a) non-assortative and (b) assortative networks of size $N=1000$ with constant degree $k=50$. For the reduced description on which our determination of $B$ is based we use $\tilde{N}=20$ and $\omega_{\text{max}},-\omega_{\text{min}}=3.126$.} \label{fig1}
\end{figure}

We next demonstrate the effect of frequency assortativity by presenting results from numerical simulations. Considering a network of size $N=1000$ with constant degree $k=50$, we generate a non-assortative network and an assortative network using $c=0$ and $c=1$, respectively, and set $\Delta_{\omega_0}=0.05$. We next solve Eq.~(\ref{eq:Kuramoto}) for each network, increasing $K$ from zero by an increment of $10^{-6}$ at each timestep $\Delta t=0.002$. Defining the order parameter
\begin{align}
R(t)=\frac{1}{Nk}\left|\sum_{i=1}^NR_i(t)\right|,\label{eq:Model02}
\end{align}
where $R_i(t)=\sum_{j=1}^NA_{ij}e^{i\theta_j(t)}$ describes the local order parameter for oscillator $i$, we plot the evolution of $R$ vs $K$ in Fig.~\ref{fig1} for the non-assortative and assortative networks in panels (a) and (b), respectively, using a solid blue curve. While the non-assortative networks displays typical behavior, transitioning from incoherence ($R\approx0$) to coherence ($R>0$) at a finite coupling strength ($K\approx0.05$), the assortative network displays much more interesting behavior. In particular, in a range of intermediate coupling strengths ($0.04\lesssim K\lesssim0.12$) the order parameter undergoes large, irregular oscillations. We will now present a dimensionality reduction method which we will use to show that the dynamics in this interesting regime are in fact chaotic.

{\it Dimensionality reduction}. The analytical technique we now summarize represents an extension of that described in Ref.~\cite{Restrepo2014EPL}. Here we assume that the general structure described above, i.e., a network described by a single degree and a collection of target frequencies, the latter specifying the distributions $g_{\omega_{0,i}}(\omega)$ from which the natural frequencies are drawn. The network is therefore characterized by the target frequency distribution $P_{\omega_0}$, which is normalized such that $\sum_{\omega_0}P_{\omega_0}=N$. The frequency assortativity of the network is captured by the function $a_{\omega_0'\to\omega_0}$, the probability that a link exists from an oscillator with target frequency $\omega_0'$ to one with $\omega_0$. We note that the assortativity function $a_{\omega_0'\to\omega_0}$ is constrained to satisfy
\begin{align}
\sum_{\omega_0}\sum_{\omega_0'}P_{\omega_0'}a_{\omega_0'\to\omega_0}P_{\omega_0}=Nk.\label{eq:assort}
\end{align}

We proceed by considering the limit of large networks, i.e., $N\to\infty$, such that the state of the network can be described by the family of distribution functions $f_{\omega_0}(\theta,\omega,t)$, where $f_{\omega_0}(\theta,\omega,t)\mathrm{d}\theta\mathrm{d}\omega/2\pi$ is the fraction of oscillators with target frequency $\omega_0$ with phase in $[\theta,\theta+\rm{d}\theta]$ and natural frequency in $[\omega,\omega+\mathrm{d}\omega]$ at time $t$. We emphasize that each natural frequency depends on the target frequency, and since $\omega$ does not change in time we have
\begin{align}
\int_0^{2\pi}f_{\omega_0}(\theta,\omega,t)\frac{{\rm d}\theta}{2\pi} = g_{\omega_0}(\omega).\label{eq:g}
\end{align}
The interaction term in Eq.~(\ref{eq:Kuramoto}) for an oscillator $j$ can be expressed in terms of the local order parameters as $K\text{Im}(e^{-i\theta_j}R_j)$. The mean-field version of the local order parameter is $R_i(t)\to R_{\omega_{0,i}}(t)$ and is given by
\begin{align}
R_{\omega_{0}}(t)=\sum_{\omega_0'}P_{\omega_0'}a_{\omega_0'\to\omega_0}\iint f_{\omega_{0}'}(\theta,\omega,t)e^{i\theta} \frac{{\rm d}\theta}{2\pi}{\rm d}\omega.\label{eq:local}
\end{align}
Finally, by the conservation of the number of oscillators, each distribution $f_{\omega_0}$ must satisfy the {\it continuity equation}
\begin{align}
0 = \partial_t f_{\omega_0}(\theta,\omega,t) +\partial_\theta[\left(\omega+K\text{Im}[e^{-i\theta}R_{\omega_0}(t)]\right)f_{\omega_0}(\theta,\omega,t)].\label{eq:cont}
\end{align}
Together, Eqs.~(\ref{eq:g}) and (\ref{eq:cont}) give a mean-field description for the macroscopic dynamics of Eq.~(\ref{eq:Kuramoto}). (We note that in other contexts the assortativity can be formulated in terms of degrees by replacing $a_{\omega_0'\to\omega_0}$ with $a_{\bm{k}'\to\bm{k}}$~\cite{Restrepo2014EPL}, or, still more generally, $a_{\bm{k}',\omega_0'\to\bm{k},\omega_0}$.)

We now follow Refs.~\cite{Ott2008Chaos,Ott2009Chaos} where the authors showed that in the long-time limit each distribution function $f_{\omega_0}$ approaches the form
\begin{align}
f_{\omega_0}(\theta,\omega,t)&=g_{\omega_0}(\omega)\left[1+\sum_{n=1}^\infty b_{\omega_0}^n(\omega,t)e^{-in\theta}+c.c.\right],\label{eq:ansatz}
\end{align}
where $c.c.$ denotes the complex conjugate of the preceding term. [Note that, since (\ref{eq:ansatz}) is the {\it time asymptotic} form of the distribution, our use of (\ref{eq:ansatz}) should yield a good approximation of all the attractor dynamics, but not necessarily the transient dynamics that describes the approach to an attractor.] Substituting Eq.~(\ref{eq:ansatz}) in Eq.~(\ref{eq:cont}), we find that each $b_{\omega_0}$ satisfies
\begin{align}
\partial_t b_{\omega_0}(\omega,t)&=i\omega b_{\omega_0}(\omega,t)+\frac{K}{2}\left[R_{\omega_0}(t)-b_{\omega_0}^2(\omega,t)R_{\omega_0}^*(t)\right].\label{eq:b}
\end{align}
Next, we substitute Eq.~(\ref{eq:ansatz}) into Eq.~(\ref{eq:local}) to obtain
\begin{align}
R_{\omega_0}(t)&=\sum_{\omega_0'}P_{\omega_0'}a_{\omega_0'\to\omega_0}\int g_{\omega_0'}(\omega)b_{\omega_0'}(\omega',t)\rm{d}\omega'.\label{eq:local2}
\end{align}

Assuming that the  frequency distribution are Lorentzian as in Eq.~(\ref{eq:Model01}), Eq.~(\ref{eq:local2}) can be simplified using the Cauchy residue theorem~\cite{Freitag2009}. In particular, it can be shown that under typical conditions~\cite{Ott2008Chaos}, each $b_{\omega_0}(\omega,t)$ is analytic in the upper-half $\omega$-plane with $b_{\omega_0}\to0$ as $|\omega|\to\infty$, which allows us to evaluate Eq.~(\ref{eq:local2}) and obtain
\begin{align}
R_{\omega_0}(t)=\sum_{\omega_0'}P_{\omega_0'}a_{\omega_0'\to\omega_0}\hat{b}_{\omega_0'}(t),\label{eq:prefinal}
\end{align}
where $\hat{b}_{\omega_0}(t)=b_{\omega_0}(\omega,t)|_{\omega=\omega_0+i\Delta_{\omega_0}}$. By setting $\omega=\omega_0+i\Delta_{\omega_0}$ in Eq.~(\ref{eq:b}), we finally obtain
\begin{align}
\frac{{\rm d}\hat{b}_{\omega_0}}{{\rm d}t}&=(i\omega_0-\Delta_{\omega_0})\hat{b}_{\omega_0}+\frac{K}{2}\left[R_{\omega_0}-\hat{b}^2_{\omega_0}R^*_{\omega_0}\right].
\label{eq:final}
\end{align}

Equations~(\ref{eq:prefinal}) and (\ref{eq:final}) govern the dynamics of a mean-field version of the full system. Importantly, this formalism can be used to reduce the dimensionality of the system. For example, Ref.~\cite{Restrepo2014EPL} dealt with the effects of degree assortativity in the absence of frequency assortativity and used an equation analogous to (\ref{eq:final}) to achieve dimensionality reduction (i.e., $\Delta_{\omega_0}\to\Delta_{\bm{k}}$, $\hat{b}_{\omega_0}\to\hat{b}_{\bm{k}}$, $P_{\omega_0}\to P_{\bm{k}}$, and $a_{\omega_0'\to\omega_0}\to a_{\bm{k}'\to\bm{k}}$). Here we use Eq.~(\ref{eq:final}) to investigate the effects of frequency assortativity in the network model described above, which has constant node degrees. We then use (\ref{eq:final}) to achieve dimensionality reduction from the original $N$ differential equations [Eq.~(\ref{eq:Kuramoto})] to a much smaller number $\tilde{N}$, by dividing the interval $[\omega_{\text{min}},\omega_{\text{max}}]$ into $\tilde{N}$ bins of width $(\omega_{\text{max}}-\omega_{\text{min}})/\tilde{N}$, where the center frequency of the $l^{\text{th}}$ bin is $\omega_0=\omega_l$, and $\omega_{\text{min}}$ and $\omega_{\text{max}}$ are chosen so that $\int_{\omega_{\text{min}}}^{\omega_{\text{max}}}[\sum_{\omega_0}P_{\omega_0}g_{\omega_0}(\omega-\omega_0)/N]{\rm d}\omega$ is nearly one. Replacing the quantity $\hat{b}_{\omega_0}$ in (\ref{eq:final}) by $\hat{b}_l$ ($l=1,\dots,\tilde{N}$) and regarding $\hat{b}_l$ as representing the collective dynamics associated with oscillators whose target frequencies fall in bin $l$, we achieve our dimensionality reduction. As we will see, $\tilde{N}$ can be made much smaller than $N$, thus greatly reducing the computational complexity. To evaluate the degree of synchronization in the reduced system, we use the order parameter
\begin{align}
B(t)=\frac{1}{N k}\left|\sum_{\omega_0,\omega_0'}P_{\omega_0}P_{\omega_0'}a_{\omega_0'\to\omega_0}\hat{b}_{\omega_0'}(t)\right|,\label{eq:R}
\end{align}
which is the reduced-system analog to the order parameter defined in Eq.~(\ref{eq:Model02}). Finally, we note that the distribution $P_{\omega_0}$ and assortativity function $a_{\omega_0'\to\omega_0}$ can be either constructed to represent an ensemble of networks or sampled from a particular network realization, as we do here.

\begin{figure}[t]
\centering
\epsfig{file =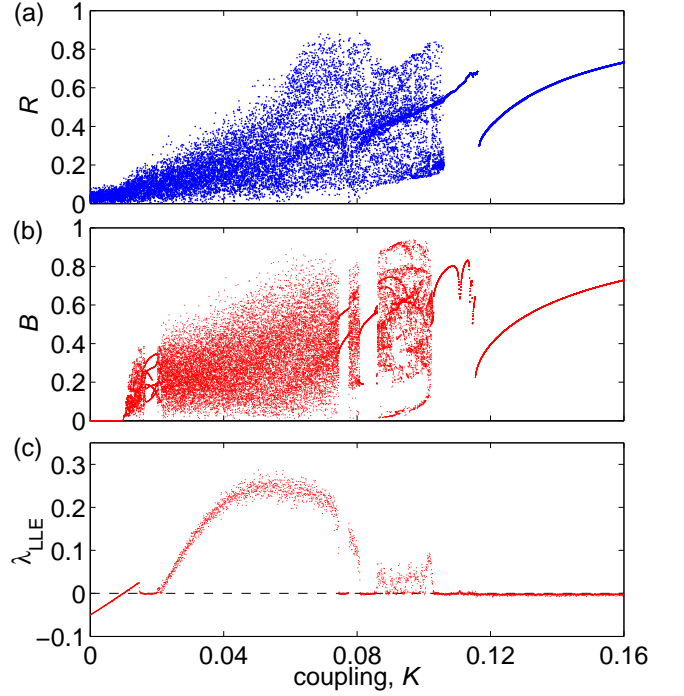, clip =,width=\linewidth } 
\caption{(Color online) \textit{Bifurcation diagrams and Lyapunov exponent.} Bifurcation diagrams of the (a) full and (b) reduced dynamics calculated by plotting the values of $R(t)$ [for (a)] and $B(t)$ [for (b)] evaluated at the times of surfaces of section piercing. (c) The largest Lyapunov exponent $\lambda_{\text{LLE}}$ as a function of $K$ calculated using the reduced system.} \label{fig2}
\end{figure}

\begin{figure*}[t]
\centering
\epsfig{file =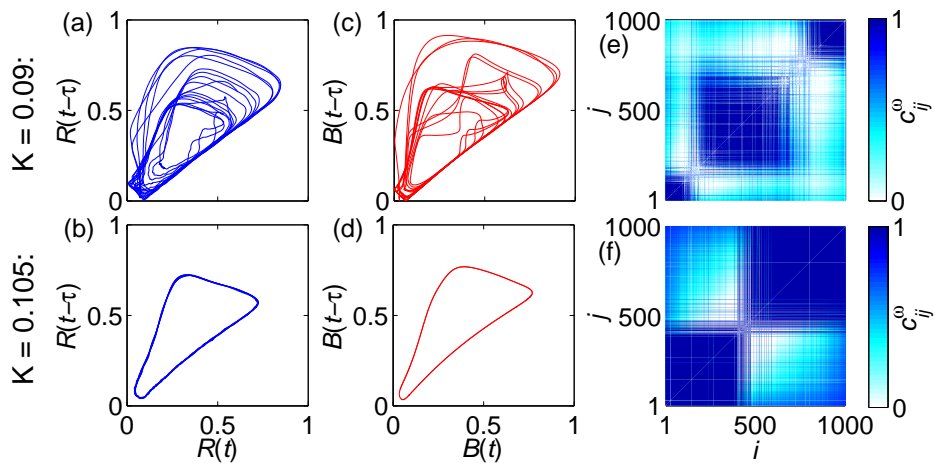, clip =,width=0.70\linewidth } 
\caption{(Color online) \textit{Chaos vs periodicity.} Time delay embeddings of the full system $[R(t),R(t-\tau)]$ for (a) $K=0.09$ and (b) $0.105$ and the reduced system $[B(t),B(t-\tau)]$ for (c) $K=0.09$ and (d) $0.105$ for $\tau=1$. Also for $K=0.09$ and $0.105$, the dynamic correlations $c_{ij}^\omega$ [(e) and (f), respectively] as calculated from the full system.} \label{fig3}
\end{figure*}

Returning to the networks obtained by the model described above, we construct the corresponding reduced systems using $\tilde{N}=20$ -- a number small enough to significantly reduce the computational cost, but large enough to retain the dynamical complexity. Solving Eq.~(\ref{eq:final}) as $K$ is increased from zero as in the full system, we plot $B$ vs $K$ for the non-assortative and assortative networks in Figs.~\ref{fig1}(a) and (b), respectively, using dashed red curves. We note that there is good agreement with the full system in both cases, and the reduced dynamics do a particularly good job of reproducing the irregular oscillations of the assortative network. Note that for small $K$ the solid blue curve in Fig.~\ref{fig1}(a) undergoes small fluctuations not present in the reduced mean-field solution (red dashed curve). These fluctuations become smaller (not shown) as $N$ is increased keeping $k/N$ fixed and can thus be explained as being due to finite network size~\cite{Hildebrand2007PRL}. The irregular oscillations for $K\lesssim0.12$ in Fig.~\ref{fig1}(b) turn out to be indicative of macroscopic chaos, as we will discuss below.

{\it Numerical investigations of chaos.} We begin by constructing bifurcation diagrams of both the full and reduced system for the assortative case. To do so, we consider the time-delay embeddings $(x,y) = [R(t),R(t-\tau)]$ and $[B(t),B(t-\tau)]$. For a given value of $K$, we record all the values of $x$ when the line $x=y$ is traversed from $y>x$ to $y<x$ after discarding transients. We use a value of $\tau=0.2$, which is large enough to overcome small finite size fluctuations present in the full system, and small enough to capture the macroscopic features of the dynamics. We present the results in Fig.~\ref{fig2}, plotting the bifurcation diagram of the full and reduced systems in panels (a) and (b), respectively. Overall the results agree well, both indicating complex oscillations and intricate behavior leading up to transitions to periodic and then stationary behavior. We note that the reduced model has thin regions of periodic behavior that we do not observe in the full system. We believe that this difference between Figs.~\ref{fig2}(a) and \ref{fig2}(b) is due to the finite size induced noise-like fluctuations present in the real network but not in the reduced network (e.g., as also present for small $K$ in Fig.~\ref{fig1}) and that this noise destroys the windows of periodicity seen in Fig.~\ref{fig2}(b) (see \cite{Barlev2011Chaos} for a related finite network size noise phenomenon). In order to test this, we first add noise to the right-hand side of (\ref{eq:prefinal}), which we then insert into (\ref{eq:final}). Simulations of this noisy model (not shown) confirm that even rather small noise is sufficient to destroy the thin regions of periodic behavior, while making a negligible effect on the dynamics for $K\gtrsim0.12$. Next, we take advantage of the lower complexity of the reduced system to calculate the largest Lyapunov exponent $\lambda_{LLE}$~\cite{Ott2002}, and plot the results in Fig.~\ref{fig2}(c). The largest Lyapunov exponent indicates that the system quickly transitions to chaos at a small coupling strength, and then intermittently transitions between chaotic ($\lambda_{LLE}>0$) and period ($\lambda_{LLE}=0$) behavior. We also investigated the behavior of the Lyapunov dimension $D_L$ by computing the whole Lyapunov spectrum (ordered $\lambda_{LLE}=\lambda_1\ge\lambda_2\ge\dots$), giving $D_L=k+(\lambda_1+\dots+\lambda_k)/|\lambda_{k+1}|$, where $k$ is the largest index such that $\lambda_1+\dots+\lambda_k>0$~\cite{Kaplan1979}. We find that $D_L$ is large near the middle of the chaotic regime and significantly decreases as $K$ is increased to approach the periodic regime (e.g., $D_L\approx13.64$ and $3.71$ at $K=0.05$ and $0.095$, respectively).

Finally, we investigate the genesis of chaotic dynamics in assortative networks. To visualize and study the dynamics we consider time-delay embeddings and frequency correlations between pairs of oscillators, defined as $c_{ij}^\omega=(1-|\omega_i^{\text{eff}}-\omega_j^{\text{eff}}|/|\omega_i-\omega_j|)^2$, where we denote the effective frequency of oscillator $i$ as $\omega_i^{\text{eff}}=T^{-1}\int_{t_0}^{t_0+T}\dot{\theta}_i(t){\rm d}t$~\cite{GomezGardenes2011PRL} for large enough $t_0$ and $T$. In particular, $c_{ij}^\omega$ quantifies the degree to which oscillators $i$ and $j$ evolve on their own ($c_{ij}^\omega=0$) or in unison ($c_{ij}^\omega=1$). We choose examples of chaotic and periodic dynamics that occur at $K=0.09$ and $0.105$, respectively, and plot in Fig.~\ref{fig3} the time-delay embeddings using $\tau=1$ for the full system [left column, panels (a) and (b)] and for the reduced system [middle column, panels (c) and (d)]. Finally, we plot the frequency correlations calculated from the full systems in the right column for both $K=0.9$ (e) and $K=0.105$ (f), with $c_{ij}^\omega=0$ and $1$ corresponding to white and blue, respectively. The correlations are plotted so that the indices $i,j$ increase with each oscillator's target frequency. First, we note that the time-delay embeddings of the full [Figs.~\ref{fig3}(a) and \ref{fig3}(b)] and reduced [Figs.~\ref{fig3}(c) and \ref{fig3}(d)] dynamics match extremely well for both the chaotic and periodic examples. Second, using the dynamic correlations we observe the formation of three [Figs.~\ref{fig3}(e)] and two [Figs.~\ref{fig3}(f)] large groups. We view such groups as meta-oscillators, and we interpret the observed dynamics as resulting from interactions of these meta oscillators. When two (one) such meta-oscillators are present the macroscopic dynamics of the order parameter is observed to be periodic (steady), while chaos can (and typically does) occur when there are three or more groups.

{\it Discussion.} In this Rapid Communication we have studied the synchronization of assortative coupled oscillator networks. Our main results are twofold. First, we have studied frequency assortativity and found that this effect can induce large and robust regions of chaotic dynamics. We have supported our results using numerical simulations of regular graphs with constant degree in order to emphasize the importance of frequency assortativity. Second, we showed that the dimensionality reduction method first presented in Ref.~\cite{Ott2008Chaos} can be extended to study this interesting case. We emphasize the strong correspondence between the dynamics of the full system and its low dimensional system reduction. In both contexts we have investigated the complicated dynamics that emerge using a combination of bifurcation diagrams, Lyapunov exponents, and time-delay embeddings. Finally, we discussed the genesis of chaos and showed that several locally synchronized groups; ``meta-oscillators,'' emerge in assortative networks, allowing for chaos. 

Chaos in the macroscopic dynamics of networks of coupled phase oscillators has been observed previously, but in different contexts. These situation include one-way coupling between two groups of coupled oscillators~\cite{Alonso2011Chaos}, globally-coupled oscillators with bimodal frequencies and an oscillating coupling strength oscillated in time~\cite{So2011Chaos}, and interacting communities of oscillators with different natural frequencies~\cite{Komarov2013PRL}. Our results show that in very simple coupled oscillator networks with fixed parameters and no external driving, chaos can be induced merely by frequency assortativity. We attribute these chaotic dynamics to the formation of three or more meta-oscillators, which is in contrast to the periodic (often called standing-wave) behavior that emerges as the result of two meta-oscillators~\cite{Restrepo2014EPL,Crawford1994}.

\acknowledgements
This work was supported by the James S. McDonnell Foundation (PSS) and ARO Grant W911NF1210101 (EO).

\bibliographystyle{plain}

\end{document}